\documentclass[aps,pre,reprint,groupedaddress]{revtex4-1}
\usepackage{graphicx}
\usepackage{color}
\usepackage{subfigure}

\begin{document}


\title{Resuspension threshold of a granular bed by localized heating}


\author{C. Morize$^{1}$}
\author{E. Herbert$^{1,2}$}
\author{A. Sauret$^{1,3}$}
\affiliation{$^1$ Laboratoire FAST, CNRS, Univ. Paris-Sud,
Universit\'e Paris-Saclay, 91405 Orsay, France}
\affiliation{$^2$ Universit\'e Paris Diderot - LIED - UMR 8236, Laboratoire
Interdisciplinaire des Energies de Demain, Paris, France}
 \affiliation{$^3$ Surface du Verre et Interfaces, UMR 125 CNRS/Saint-Gobain 39, quai
Lucien Lefranc, F-93303 Aubervilliers Cedex, France}

\date{September, 5 2017}

\begin{abstract}
The resuspension and dispersion of particles occur in industrial fluid dynamic processes as well as environmental and geophysical situations. In this paper, we experimentally investigate the ability to fluidize a granular bed with a vertical gradient of temperature. Using laboratory experiments with a localized heat source, we observe a large entrainment of particles into the fluid volume beyond a threshold temperature. The buoyancy-driven fluidized bed then leads to the transport of solid particles through the generation of particle-laden plumes. We show that the destabilization process is driven by the thermal conductivity inside the granular bed and demonstrate that the threshold temperature depends on the thickness of the granular bed and the buoyancy number, i.e., the ratio of the stabilizing density contrast to the destabilizing thermal density contrast.
\end{abstract}


\maketitle

\section{Introduction}

The transport of solid particles induced by shearing particulate beds with water or air flow occurs in many geophysical events, such as in river beds and landscape evolution, wind-blown sand, and dust emission \cite{seminara2010fluvial,charru2013sand}, but also in industrial processes: filtration systems or the food industry, for instance \cite{ruth1933studies,wyss2006mechanism,dressaire2017clogging,werner2007air,iveson2001nucleation}. Depending on the nature, geometry, and regime of the fluid flow, different types of particle movements can be observed. Conventionally, rolling and sliding motion, saltation, and suspension are distinguished depending on the Reynolds number. In other configurations, when the granular bed is subject to an ascending flow of liquid or gas, the normal stress can fluidize the particulate medium and maintain grains in suspension, which leads to particular properties and characteristics of the medium \cite[see e.g.,][]{deen2007review}. This situation is observed for instance during the rise of air in an immersed granular bed \cite{varas2011venting,ramos2015gas}.

In addition of these mechanical mechanisms of resuspen- sion, a heat source inside or below the sediment may destabi- lize a loose random packing of granular matter. Resuspension due to thermal effects is of great importance to understand, for example, volcanic ash clouds \cite{carazzo2013particle} or seafloor hydrothermal systems such as black smokers. Whereas the resuspension and fluidization of an immersed granular bed by fluid flows such as vortices \cite{munro2009sediment,bethke2012resuspension,yoshida2012collision,masuda2012collision},
impacting jets \cite{badr2014erosion,sutherland2014bedload}, shear
flows \cite{charru2004erosion,hong2015onset}, or gas crossing a liquid-saturated granular bed have been the focus of many studies, the ability of thermal convection to resuspend particles remains poorly understood. Indeed, in recent decades particular attention has been focused on the settling of particles, initially in suspension, in steady cellular convection but only few studies \cite{solomatov1993entrainment} have investigated the destabilization process of an initially loose randomly packed granular bed driven by thermal convection.

Several scenarios are possible to induce the reentrainment of solid particles from the bottom: erosion by the bottom shear stress induced by convection current or fluidization by emergence of particle-laden plumes. To explore the mechanisms of the
resuspension process by thermal convection, Solomatov \emph{et
al.} \cite{solomatov1993entrainment,solomatov1993suspension} used a three-dimensional experimental setup with aqueous solution and polystyrene particles which are initially sedimented to form a loose random packing of granular matter. The bottom wall is uniformly heated from below to ensure the destabilization process of the sedimented particles. The authors defined a Shields number, Sh, which compares the ratio of the destabilizing hydrodynamic stress $\tau$
exerted on a grain to its stabilizing apparent weight $\Delta \rho \,
g \,d$, where $\Delta \rho = \rho_p-\rho_l$ is the density difference between the grains and the ambient fluid, $g$ is the acceleration of the gravity and $d$ is the mean particle diameter.
The resuspension of particles from the bottom is driven by the tangential buoyancy stress $\tau = \beta\, \rho_l \,g \,\Delta T \,
\delta_T$ (where $\beta$ is the thermal expansion coefficient,
$\Delta T$ is the temperature difference between the bottom and
the top of the cell and $\delta_T$ is the thermal boundary layer
thickness) at the top of the granular bed induced by convection current: particles roll and slide horizontally and can form dunes. At the crest of a dune, the tangential buoyancy stress is vertical and can lead to the resuspension of particles. The buoyancy Shields number, separating a regime without bed motion from a regime with particle entrainment is approximatively constant and of the order of 0.1 in the experiments. On the other hand, Martin and Nokes \cite{martin1988crystal} used a similar setup but a different initial state where particles are in suspension. The reentrainment of particles occurred by the emergence of plumes from the bottom but the authors were unable to extract a criterion for reentrainment.

The aim of the present work is to investigate the resus- pension of a granular bed of particles by fluidization from a localized heat source. This situation is different from the uniform heating used in previous studies \cite{solomatov1993entrainment} where the gran- ular bed was eroded over a long time by shear flows induced by the thermal effects and the particles were susceptible to rolling, sliding, and being carried away at the bed surface by saltation \cite{solomatov1993suspension}. Here, we demonstrate that at larger temperature difference the fluidization of the granular bed can also occur from the bottom and actively participate in the resuspension and reentrainment of particles into the bulk through the generation of particle-laden plumes. We focus on the scaling of this fluidization threshold with the relevant dimensionless parameters that describe this destabilization process.

\section{Experimental methods}

\begin{figure}
\includegraphics[width=0.45\textwidth]{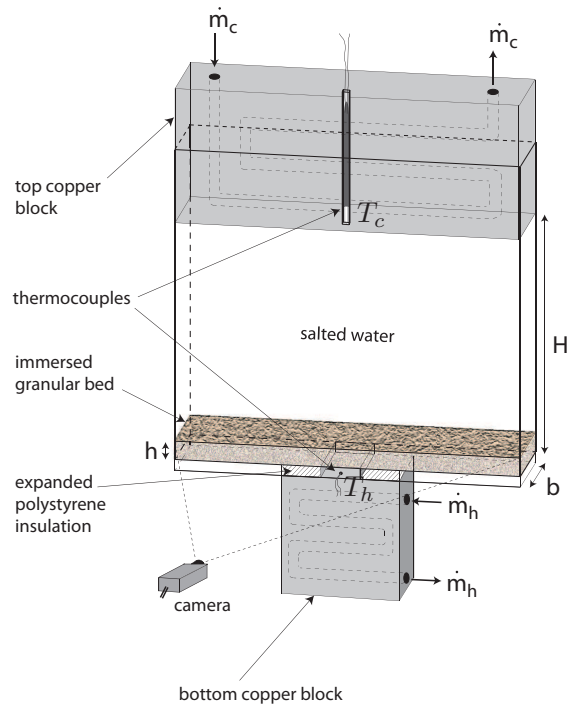}
\caption{Schematic of the experimental
setup.\label{Fig1_ExpSetup}}
\end{figure}

We study the resuspension of spherical polystyrene particles (Dynoseeds purchased from MicroBeads) of diameter $2\,a=250\,\pm 10 \mu{\rm m}$ and a density $\rho_p^0=1.049 \pm 0.003 \,{\rm g\,cm^{-3}}$ \cite{boyer2011suspensions} induced by a localized heating at the bottom of the granular bed. The setup used in the resuspension experiments is shown in Fig. \ref{Fig1_ExpSetup} and consists of a rectangular Poly(methyl methacrylate) tank of internal dimensions $20$ cm in length, $b=1.2$ cm in width and $H=20$ cm in height. Given the aspect ratio of the tank, no significant motion of the fluid takes place in the short direction. The temperature is imposed at the top and the bottom of the tank through two copper plates whose temperatures are imposed by two circulating thermostated baths and measured by platinum thermocouples located inside the plates. The top copper plate is 20 cm wide and covers the tank, whereas the bottom copper plate is centered and has a width of 4 cm, resulting in a localized heating.

The working fluid is composed of water with different concentrations of calcium chloride, CaCl$_2$. The quantity of salt is varied to increase the density of the working fluid in the range $\rho_l=[1,\,1.049]\,{\rm g.cm^{-3}}$ (measured at 20$^{\rm
o}$C with a densimeter Anton Paar DMA 35;  see Appendix). In addition, a small amount of sodium dodecyl sulfate surfactant is added to the water to initially disperse the dry particles into water and avoid the trapping of air bubbles. Each experiment is performed with a new suspension of particles that is allowed to settle and form a loose randomly packed granular bed at the bottom of the tank. The resulting compacity of the granular bed, i.e., the solid volume fraction $\phi=V_p/V_{tot}$ (where $V_p$ is the volume of the particles and $V_{tot}$ is the volume of the liquid and the particles) is equal to $\phi=0.57 \pm 0.01$ and its controlled thickness $h$ lies in the range $1$ to $30\,{\rm mm}$. A vertical laser sheet is set through the short side of the tank so that the fluid and the granular bed are clearly visible. The evolution of the granular bed is recorded using a Phantom MIRO M110 camera
(resolution $1200\times 800$ pixels) at 1 frame/s.

At the beginning of each experiment, the thermostated baths are first allowed to reach their assigned temperature, leading to a top temperature $T_c$ and a bottom temperature in the copper plate, $T_h$. Initially, we set $T_c=T_h=T_0=15^{\rm o}$C. We then wait for a sufficiently long time, typically 30 min, to ensure that the initial temperature in the system is homogeneous and equal to $T_0=15^{\rm o}$C. At time $t=0$, we suddenly increase the temperature of the bottom copper plate to the reference temperature
$T_{h}>T_0$. Experimentally, providing a sudden increase in temperature of the bottom plate is challenging and we rely on a third thermostated bath that is set at $\theta_{h}$ prior to the experiment. Switching the fluid flow in the bottom plate to this thermostated bath leads to a progressive increase of the temperature of the bottom copper plate, $T_h$, while the upper copper plate remains at $T_c=T_0=15\,^{\rm o}$C. The temperature $T_h$ increases continuously and, when it reaches a critical value, we observe the sudden destabilization and resuspension of the granular bed.

\section{Experimental results}

\subsection{Phenomenology}

 \begin{figure}
 \includegraphics[width=0.45\textwidth]{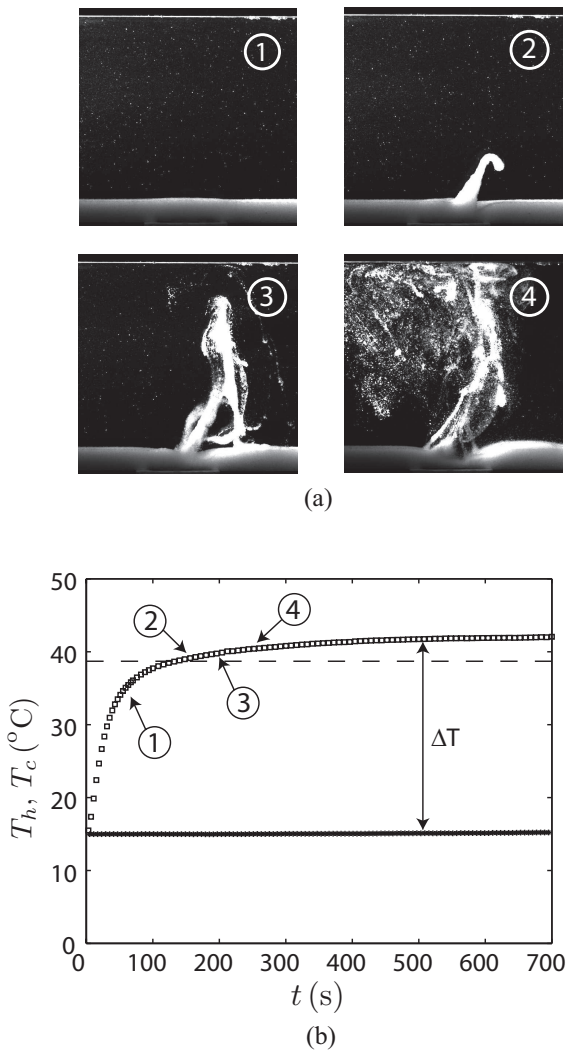}%
 \caption{(a) Time lapse of an experiment showing the granular bed (1) before and (2) after the resuspension threshold and (3), (4) the later evolution of the particle-laden plume. The experimental parameters are $h=10$ mm, the temperature of the thermostated bath $\theta_h=65^{\rm o}\rm{C}$, and $\Delta \rho^0=1.98\,{\rm kg\,m^{-3}}$. (b) (b) Time evolution of the temperature in the top, $T_c$ (crosses), and the bottom, $T_h$ (squares), , of the cell as a function of time. The horizontal dotted line indicates the resuspension threshold. The experimental pictures (1)Ð(4) are also indicated in the plot.\label{Fig2_Phenomenology}}
 \end{figure}

A typical experiment is shown in Fig. \ref{Fig2_Phenomenology}(a) where a time lapse shows the evolution of the granular bed. Initially, the temperature profile is constant in the granular and the fluid layers, equal to $T_0=15^{\rm o}{\rm C}$, and the resulting density profile of the liquid is constant in both regions and is equal to $\rho_{l}^0$. The density of the polystyrene beads at $T_0$, $\rho_p^0$, is larger that the density of the liquid, $\rho_{l}^0$, and therefore the granular bed is stable. At time $t=0$, the bottom copper plate is connected to the hot thermostated bath at the temperature $\theta_{h}$, and the temperature $T_h$ in the bottom plate then starts to increase while the temperature $T_c$ of the top plate remains constant as shown in Fig. \ref{Fig2_Phenomenology}(b). The progressive increase of the temperature of the hot source, $T_h$, leads to the increase of the temperature in the bottom region of the cell and therefore a decrease of the liquid and particle densities in the granular bed. Because of the thermal loss between the thermostated bath, whose temperature $\theta_h$ is fixed, and the bottom copper plate, the final temperature $T_h^{f}$ is smaller than $\theta_h$. Indeed, the thermostated bath is connected to the bottom copper plate through tubings, which, even insulated, lead to heat transfer with the ambient atmosphere and lead to a temperature $T_h$ smaller than $\theta_h$. Nevertheless, the temperature probe measures $T_h$ during the entire duration of an experiment and we therefore refer to this temperature in the following.

During the first phase, no motion of the granular bed occurs (corresponding to region 1 in Fig. \ref{Fig2_Phenomenology}). Then, at a temperature threshold $T_h^*$, the granular bed starts to destabilize with the formation of a small corrugation that later grows in time (regions 2 and 3). The transition from a flat granular bed to the rise of a particle-laden plume occurs suddenly over a time scale that is small compared to the duration of the entire experiment and the evolution of the temperature at the bottom of the cell. We can therefore accurately estimate the time and the corresponding temperature $T_h$ at the bottom plate at which the granular bed becomes unstable, which in this example corresponds to $t=135\,{\rm s}$ and $T_h^* \simeq 38^{\rm o}{\rm C}$.

In the following, we rationalize quantitatively the threshold temperature $T_h^*$ and the destabilization time at which the granular bed starts to fluidize. We systematically investigate the influence of the temperature of the hot source, $T_h$, the initial density contrast between the fluid and the particles, $\Delta \rho^0$, and the thickness of the granular bed, $h$.

\subsection{Time-evolution of $T_h$} \label{sec:T}

  \begin{figure}
 \includegraphics[width=0.5\textwidth]{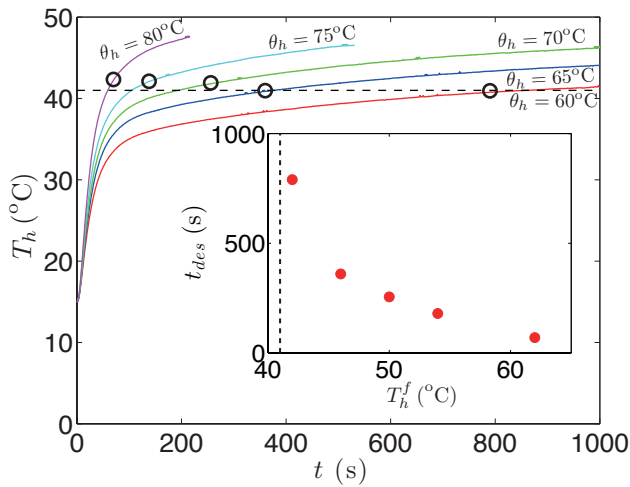}%
 \caption{Time evolution of the temperature $T_h$ at the bottom copper
 plate for different temperature of the hot thermostated bath: $\theta_h=60^{\rm o}{\rm C}$ (red), $65^{\rm o}{\rm C}$ (blue), $70^{\rm o}{\rm C}$ (green), $75^{\rm o}{\rm C}$ (cyan) and $80^{\rm o}{\rm C}$  (magenta). The open black circles indicates the threshold temperature $T_h^*$ and the corresponding destabilization time $t_{des}$ in each situation. The horizontal black dotted line indicates the value of $T_h^*$ for a slow increase of temperature.
 In these experiments, $h=10\,{\rm mm}$ and $\Delta \rho^0= 2.6 \,\rm{kg\,m^{-3}}$.
 Inset: Time $t_{des}$ needed to reach the threshold in temperature as a function of the steady temperature of the bottom plate $T_h^f$. The vertical black dotted line indicates $T_h^*$.
 \label{Fig3a_Quantitative}}
 \end{figure}

We observe that a threshold temperature $T_h^*$ needs to be reached to
resuspend locally the granular bed. Because our experimental approach involves a transient increase of
the temperature of the bottom copper plate, we characterize
the influence of this transient dynamic on the threshold
temperature. To do so, we consider a granular bed of fixed
height $h=10\,{\rm mm}$ and a
fixed initial density contrast $\Delta \rho^0=\rho_p^0-\rho_l^0=
2.6 \,\rm{kg\,m^{-3}}$. We then impose different temperatures at
the thermostated bath connected to the bottom hot source,
$\theta_h=[60,\,65,\,70,\,75,\,80]\,^{\rm o}{\rm C}$, keeping the
temperature at the top of the cell constant and equal to
$T_c=15\,^{\rm o}C$ during the entire experiment. As mentioned previously, the thermal loss between the thermostated bath and the bottom copper plate leads to a temperature $T_h^{f}$ reached at the bottom of the cell significantly smaller than the temperature $\theta_h$ of the thermostated bath. Experimentally, we determine that $T_h^{f}$ is in
the range $42\,^{\rm o}$C to $64\,^{\rm o}{\rm C}$.

The time evolution of the temperature at the bottom plate, as recorded by the temperature probe, is shown in Fig. \ref{Fig3a_Quantitative} for varying values of the temperature $\theta_h$ of the hot thermalized bath. The threshold temperature $T_h^*$ at which the granular bed starts to resuspend is also reported. We observe that the temperature of the bottom plate increases quickly at the beginning and then the increase in temperature becomes slower. In all the different situations considered here where $T_h^f > T_h^*$, we
observe the resuspension of the granular bed at sufficiently long
time. In addition, the threshold temperature remains
approximatively constant, $T_h^*=41.6\pm0.8\,^{\rm o}{\rm C}$,
which suggests that, for a constant thickness of granular layer sufficiently large,
here $h=10$ mm, and given density contrast, the threshold temperature $T_h^*$
does not depend significantly on the dynamics of the system. However, we are going to see in the following that the transient effects are required to describe the system when varying the granular bed thickness $h$.

Although the threshold temperature for the destabilization of the granular bed does not seem to depend significantly on the imposed value of the temperature of the hot thermostated bath $\theta_h$, the time needed to reach the destabilization threshold, $T_h^*$,
increases sharply when $\theta_h$ is decreased. This observation is mainly explained by the thermodynamics of the system: the time needed to reach the threshold temperature increases when decreasing the temperature of the hot source, leading to a longer waiting time. In addition, if the temperature of the bottom plate remains smaller than $T_h^*$, no destabilization of the granular bed is observed at long time. In the following, we consider a heat flux at the hot source that leads to a maximum steady value of the bottom plate of $T_h^f \simeq 50\,^{\rm o}{\rm C}$.

\subsection{Density contrast between the fluid and the particles $\Delta \rho^0$} \label{sec:rho}

  \begin{figure}
 \includegraphics[width=0.5\textwidth]{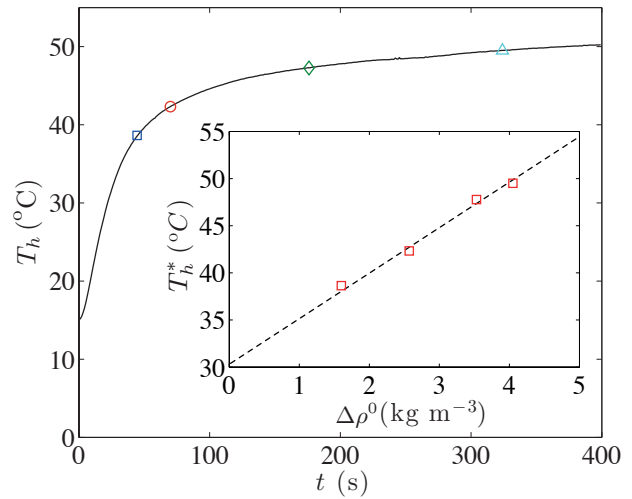}%
 \caption{Time evolution of the bottom plate temperature $T_h$ (continuous line) and
 threshold values (open symbols) for various density contrast $\Delta \rho^0 =
 \rho_p^0-\rho_l^0$ (blue square, $\Delta \rho^0 = 1.6\,{\rm kg\,m^{-3}}$; red circle, $\Delta  \rho^0 = 2.6\,{\rm kg\,m^{-3}}$; green diamond, $\Delta \rho^0 = 3.5\,{\rm kg\,m^{-3}}$;  cyan triangle, $\Delta \rho^0 = 4.0\,{\rm kg\,m^{-3}}$). Inset: Threshold temperature $T_h^*$  as a function of $\Delta \rho^0$. The dotted line is a linear fit. The experimental parameters are $h=10$ mm and the temperature of the hot thermostated bath is set at $65^{\rm o}\rm{C}$.
 \label{Fig3b_Quantitative}}
 \end{figure}

Another relevant parameter in the destabilization process is the density contrast between the fluid and the particles, $\Delta \rho^0$. We thus vary the density of the fluid, $\rho_l^0$, by tuning the concentration of CaCl$_2$ salt so that $\Delta \rho^0 = \rho_p^0-\rho_l^0$ ranges from $1.6\,{\rm kg\,m^{-3}}$ to $4.1\,{\rm kg\,m^{-3}}$ (see Appendix). No resuspension of the granular bed has been observed with further increase of the density difference $\Delta \rho^0$ within the range of temperatures that we have access to in our experiments. For a constant bed thickness $h=10\,{\rm mm}$ and $\theta_h = 65\,^{\rm o}{\rm C}$, we report in Fig. \ref{Fig3b_Quantitative} that the resuspension of the granular bed occurs at different temperature thresholds. Indeed, the smaller the density contrast $\Delta \rho^0$ is, the sooner the resuspension takes place during the temperature ramp, which corresponds to smaller temperature threshold of the bottom plate $T_h^*$. The inset of Fig. \ref{Fig3b_Quantitative} highlights that the temperature threshold $T_h^*$ increases linearly with the density contrast $\Delta \rho^0$
 in the range of parameters that we have considered. We come back to this point in the next section when we rationalize our results with dimensionless parameters.

  \subsection{Granular bed thickness $h$} \label{sec:h}

\begin{figure}
 \includegraphics[width=0.475\textwidth]{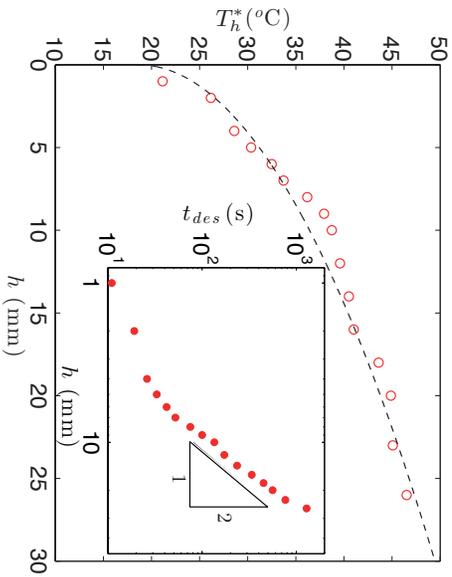}%
 \caption{Temperature threshold $T_h^*$ for increasing thickness of the granular bed, $h$. Red circles are the experimental results and the black dotted line shows $T_h^* \propto h^{1/2}$. Inset: Time elapsed, $t_{des}$, prior to the resuspension threshold for increasing thickness of the granular bed $h$. The temperature of the hot thermostated bath is set at $\theta_h=65^{\rm o}\rm{C}$ and $\Delta \rho^0=1.98\,{\rm kg\,m^{-3}}$. \label{Fig5_Thickness}}
 \end{figure}

Finally, we investigate the influence of the granular bed thickness on the temperature threshold and report the results in Fig. \ref{Fig5_Thickness}. We keep the density of the liquid and the particles constant, and using the same temperature of the hot thermostated bath $\theta_h$, we measure the temperature threshold at which the resuspension occurs as well as the time elapsed before the destabilization, $t_{des}$.
We observe that the threshold temperature increases with the
thickness of the granular layer following a trend close to $T_h^*
\propto h^{1/2}$. In addition, the time $t_{des}$ follows a slope
$t_{des} \propto h^2$. This scaling suggests that the mechanism responsible for the resuspension of the granular bed is diffusive.

\section{Discussion}

The experimental results highlight that an increase in the density
contrast $\Delta \rho^0$ or the granular bed thickness $h$
leads to a larger temperature threshold $T_h^*$ for
destabilization. In this section, we show that these results can be rationalized by considering buoyancy effects in the granular bed.

\subsection{Buoyancy number}

Buoyancy-driven flows in a Hele-Shaw cell are commonly described
using a modified Rayleigh number defined as \cite{Hartline1977}:
\begin{equation}
Ra=\frac{g\,\beta\,\Delta T\,H\,b^2}{12\,\nu\,D},
\end{equation}
where $g$ is the acceleration due to gravity, $b$ and $H$ the
width and height of the cell, respectively, $\beta$ is the thermal
expansion coefficient, $\nu$ is the kinematic viscosity and $D$ is
the thermal diffusivity. For large enough Rayleigh number, the
system is unstable and natural convection is observed. In the
present study, the Rayleigh number lies in the range
$[10^3,\,10^6]$, and we observe large recirculation cells in the top layer (liquid) even below the resuspension threshold. We therefore need to consider an additional dimensionless number to explain the global destabilization of the granular bed.

  \begin{figure}
 \includegraphics[width=0.475\textwidth]{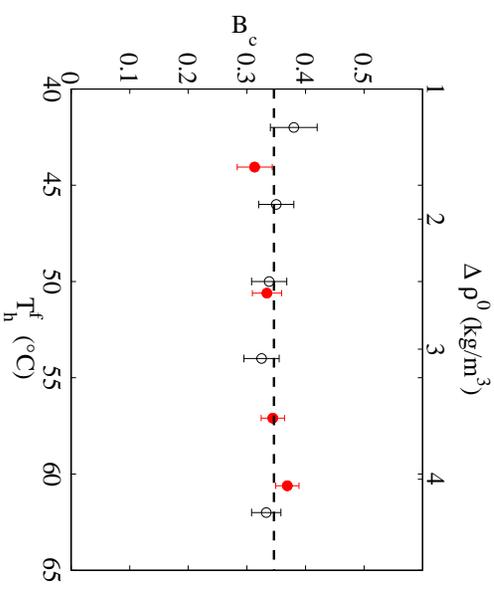}%
 \caption{Evolution of the critical buoyancy number $B_c$ varying the temperature $T_h^f$, hence $\Delta T$ (open black circles) or the density contrast $\Delta \rho^0$ (solid red circles) for a constant granular bed thickness $h=10$ mm. \label{Fig_Buoyancy}}
 \end{figure}

Here, the presence of a granular bed at the bottom of the cell leads to a more complex situation as we now observe the possible destabilization of the granular layer beyond a temperature threshold. This situation can be related to past studies that have considered a density stratification in two-layer Newtonian fluids when the two fluids have different densities and viscosities and there is no surface tension between the two fluid layers \cite{Davaille1999,LeBars2002,LeBars2004}. In this situation, it has been shown that the onset of convection can be either stationary or oscillatory depending on a dimensionless number, the buoyancy number $B$, defined as the ratio of the stabilizing
 density anomaly to the destabilizing thermal density
anomaly:
\begin{equation}
B=\frac{\rho_p^0-\rho_l^0}{\rho_l^0\,\beta\,\Delta T}.
\end{equation}
where $\Delta T=T_h-T_c$. Depending on the
buoyancy number $B$, two main regimes have been identified.
Typically, when $B$ is large enough, i.e., $B$ larger
than 0.5-1, convective flows develop above and/or below the flat interface to obtain the so-called stratified regime. When the buoyancy number $B$ is small enough, typically smaller than 0.3-0.5, the interface can become unstable and spontaneous flow occurs in the whole tank, leading to the mixing of the two initial fluid layers in the bed and above. Using this approach, we rescale the experimental results presented in Secs. \ref{sec:T} and
\ref{sec:rho}, obtained for a given thickness $h$ of the immersed
granular bed and varying the density contrast and the temperature of the
thermostated bath. The results reported in Fig. \ref{Fig_Buoyancy}
highlight that the destabilization of the granular bed occurs for
a constant value of the buoyancy number $B=B_c \simeq 0.34$ at $h=10$ mm.

\begin{figure}
 \includegraphics[width=0.475\textwidth]{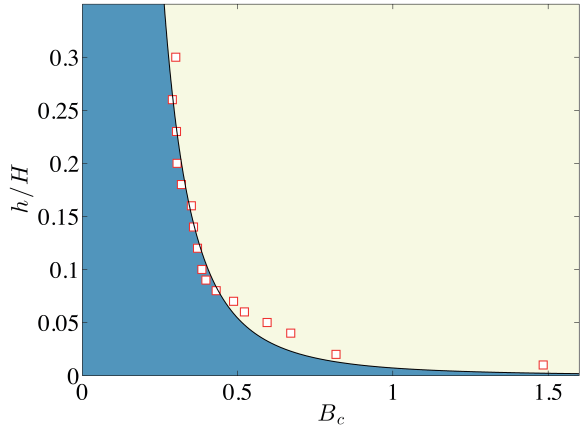}%
 \caption{Evolution of the critical buoyancy number $B_c$ varying the thickness of the granular bed, $h$. The temperature of the thermostated bath is set at $\theta_h=65^{\rm o}\rm{C}$ and $\Delta \rho^0=1.98\,{\rm kg\,m^{-3}}$. The black solid line is the best polynomial fit and is a guide for the eye. In the light yellow region, the granular bed is stable whereas resuspension occurs in the dark blue region. \label{Fig_Buoyancy2}}
 \end{figure}

In Fig. \ref{Fig_Buoyancy2}, we also report the evolution of the
critical buoyancy number $B_c$ for varying $h$ while the
other parameters are kept constant. These results show that $B_c$ decreases
when increasing the rescaled granular bed thickness $h/H$ as
observed for two Newtonian fluids by Le Bars \& Davaille \cite{LeBars2004}. However, such an approach does not explain the evolution of $t_{des}$ with the bed thickness.

 \subsection{Destabilization threshold}

We consider a granular bed of compacity $\phi$ made of polystyrene particles of
density $\rho_{p}^0$. Initially, the system is at temperature
$T_c=15^{\rm o}{\rm C}$. We consider an infinitesimal element of
length ${\rm{d}}L$ and width $b$, which is the gap of the
Hele-Shaw cell. The density of the granular bed averaged over the thickness of the
granular bed is written
\begin{equation} \label{Theo_1}
\langle \rho(T) \rangle_h = \frac{1}{h}\int_{0}^h\,\phi\,\rho_{p}(T)+({1-\phi})\,\rho_l(T)\,{\rm d}z,
\end{equation}
where the temperature $T(z)$ depends on the vertical coordinate.
The evolution of the density with the temperature is given by :
\begin{equation} \label{Theo_2}
\rho_l(z)=\rho_l^0\,\left[1-\alpha_l(T)\,\Delta T\right],
\end{equation}
for the liquid and by
\begin{equation} \label{Theo_3}
\rho_{p}(z)=\rho_p^0\,\left[1-\alpha_{p}(T)\,\Delta T\right],
\end{equation}
for the polystyrene beads. In these equation, $\alpha_l$ and $\alpha_p$ are the coefficient of thermal expansion of the liquid and polystyrene beads, respectively (see Apppendix). 

\begin{figure}
 { \includegraphics[width=0.40\textwidth]{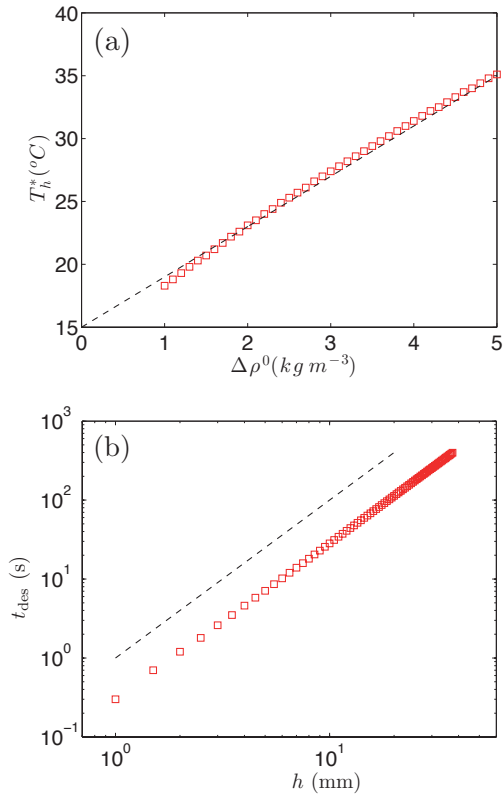}}
 \caption{(a) Threshold temperature $T_h^*$ obtained in the steady state regime as a function of the density contrast $\Delta \rho^0$. The thickness of the granular bed is $h=10\,{\rm mm}$. The dotted line shows the linear scaling $T_h^* \propto \Delta \rho^0$. (b) Time elapsed, $t_{des}$, prior to the resuspension threshold obtained by solving the diffusion equation for increasing thickness $h$ of the granular bed. The numerical parameters are $T_h=55^{\rm o}{\rm C}$, $\Delta \rho = 5\,{\rm kg\,m^{-3}}$. The dotted line is a slope $t\propto h^2$. \label{theo_drho}}
 \end{figure}

We first consider the steady state where the temperature profile only depends on the vertical coordinate $z$. Experi- mentally, we observe, even below the resuspension threshold, large-scale flow in the fluid and we therefore assume that at the top of the granular bed the local temperature is comparable to the temperature of the cold source, $T_c$. Therefore, the temperature profile in the layer can be approximated as
$T(z)=T_c+(T_h-T_c)\,z/h$. In this configuration, the granular bed becomes unstable when the density averaged over the entire thickness h becomes smaller than the density of the fluid on top of it, assumed to be at the temperature $T_c=T_0$, which means if ${\langle \rho(T) \rangle_h}< {\rho_l^0} $. We
can solve this condition numerically for varying density
contrast $\Delta \rho^0$ as illustrated in Fig. \ref{theo_drho}(a).

We observe that qualitatively the threshold temperature $T_h^*$
increases linearly with the density contrast $\Delta \rho^0$ as
observed experimentally. This observation is consistent with the
definition of the Buoyancy number $B$. For a constant granular bed thickness, the density contrast has a stabilizing effect, whereas an increase in temperature contributes to the destabilization of the granular bed through a local modification of the density of the granular bed. We should also emphasize that this description is qualitative but not quantitative because of the experimental limitations. Indeed, the increase in temperature at the bottom copper plate is not instantaneous and therefore $T_h$ increases as a function of time. Nevertheless, this approach allows us to highlight the influence of the buoyancy number on the resuspension threshold of an immersed granular bed and the destabilizing effect of the temperature.

The influence of the transient state is clearly observed when considering the influence of the granular bed thickness $h$. Indeed, if we consider the steady state only, the temperature threshold $T_h^*$ should not depend on $h$. However, our experimental results show that $T_h^*$ increases with $h$ (see Fig. \ref{Fig5_Thickness}). Although in our experiments the time variation of the temperature of the hot source is intrinsically related to the thermal properties of the system, we can consider the effect of the time variation to explain the scaling of the time to destabilize the granular bed, $t_{des}$. We modified the model developed previously and now consider that, at $t<0$, the temperature is equal to $T_c$ everywhere in the fluid and in the granular bed. At time $t=0$, the lower part at $z=0$ is suddenly put at $T=T_h$. In the experiment, this temperature is increasing when connecting the hot thermostated bath but here, for the sake of simplicity, we assume that the temperature of the bottom plate is reached instantaneously. As a result, the time dependence of the temperature at the position $z$ is the solution of the classical one-dimensional diffusion problem
\begin{equation} \label{Theo_4}
T(z)=T_c+\left(T_h-T_c\right) \left[1-{\rm erf}\left(\frac{z}{2\,\sqrt{D\,t}}\right)\right],
\end{equation}
with $D \simeq 1.8 \times 10^{-7}\,{\rm m^2.s^{-1}}$, the effective diffusion coefficient in the granular bed. We know that the granular bed becomes unstable when ${\langle \rho(T) \rangle_h}< {\rho_l^0}$, and we can solve this condition numerically using equations (\ref{Theo_1})-(\ref{Theo_4}).

The corresponding results are reported in Fig. \ref{theo_drho}(b):
the scaling observed experimentally $t_{des} \propto h^2$ is captured by the diffusion equation in the granular bed. Therefore, transient effects appear to be important to explain the destabilization of an immersed granular bed. The qualitative model presented here captures the key physical effects and provides scaling laws to describe the phenomenon. To take into account the increase in temperature of the hot source a full numerical model will be needed.

  \section{Conclusion}

In this paper, we have explored experimentally the resus- pension of an immersed granular bed by a localized heat source. The granular bed is made of particles that have a density slightly smaller than the density of the surrounding fluid. Our experiments illustrate that, beyond a temperature threshold, the averaged density of the granular bed can become smaller than the density of the top fluid and produce an overturning instability as observed for two Newtonian fluid \cite{Davaille1999,LeBars2002,LeBars2004}. This flow results in the production of thermal plume and the dispersion of the granular particles in the entire container. We have shown that this mechanism can be described through the buoyancy number
$B={(\rho_p^0-\rho_l^0)}/({\rho_l^0\,\beta\,\Delta T})$. The threshold
value $B_c$ is observed to be dependent on the granular bed thickness in our experiments owing to the transient regime. We rationalize our experimental findings with scaling argu- ments that capture the main features of this destabilization process.

Such resuspension of a granular bed could be important in geophysical and environmental processes in which localized heating can induce the transport of deposited particles and the later contamination of the environment. The dynamics of particle-laden plumes is important to describe the subsequent dispersion of the particles.

\vspace{0.5cm}

\begin{acknowledgments}
The authors are grateful to P. Gondret and A. Davaille for fruitful discussions and to E. Dressaire for a careful reading of the manuscript. We acknowledge J. Amarni, A. Aubertin, L. Auffray, C. Borget, and R. Pidoux for experimental help. This work was supported by the French ANR (project Stabingram ANR
2010-BLAN-0927-01). C.M. also acknowledges support from the AAP
``Attractivit\'e Jeune chercheur'' from Paris-Sud University.
\end{acknowledgments}

\appendix*

\section{Density of the water solution and the polystyrene beads}

\subsection{Aqueous solution}

\begin{figure}
\begin{center}
\includegraphics[width=7cm]{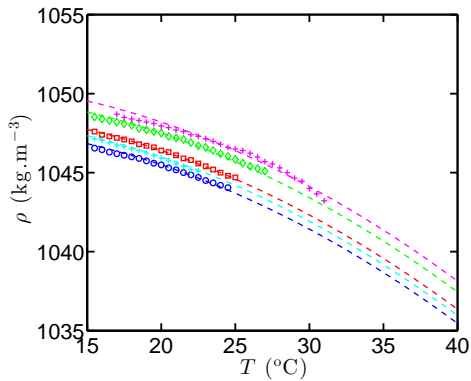}
\caption{{Density of aqueous mixture for different salt concentration: $4.85\%$ w/w (blue circles), $4.9\%$ w/w (cyan crosses), $4.95\%$ w/w (red squares), $5.05\%$ w/w (green diamonds) and $5.15\%$ w/w (magenta crosses). The dotted lines are the best fits from the equation (\ref{alpha}) for varying $\rho^0_{l}(c)$.}}\label{density}
\end{center}
\end{figure}

The liquid used in our experiments is a mixture of distilled water and CaCl$_2$ (between 0\% and 5.5\% w/w). We measured the density of different aqueous mixture using a densimeter (Anton Paar DMA 35) in a range of temperature between $15^{\rm o}$C and $30^{\rm o}$C as reported in Fig. \ref{density}.

The evolution of the density with the temperature and the salt concentration $\rho_l(c,T)$ is fitted using the expression
\begin{equation} \label{alpha}
\rho_l(c,T)=\rho^0_{l}(c)\,\left[1-\alpha(T)\,(T-T_0)\right],
\end{equation}
where $\rho^0_{l}(c)$ is the density of the aqueous mixture having a concentration $c$ of CaCl$_2$ (w/w), taken at the temperature $T_0=20^{\rm o}$C. The coefficient of thermal expansion, $\alpha(t)=a\,T+b$, is fitted from the experimental data ($a=9.6\times 10^{-6}\,(^{\rm o}{\rm C})^{-2}$, $b=1.2\times 10^{-4}\,(^{\rm o}{\rm C})^{-1}$ and $T$ is expressed in $^{\rm o}{\rm C}$).

\subsection{Polystyrene beads} 

The coefficient of thermal expansion (volume) is taken equal to $\alpha_{p}=1.9\,\times 10^{-4}\,(^{\rm o}{\rm C})^{-1}$ \cite{polystyrene}, such that the density of the polystyrene beads can be expressed as
\begin{equation} 
\rho_{p}(T)=\rho_{p}^0\,\left[1-\alpha_{p}\,(T-T_0)\right],
\end{equation}
with $\rho_{p}^0=1049 \,{\rm kg.m^{-3}}$ taken at $T_0= 20^{\rm o}$C.

%


\begin{thebibliography}{27}%
\makeatletter
\providecommand \@ifxundefined [1]{%
 \@ifx{#1\undefined}
}%
\providecommand \@ifnum [1]{%
 \ifnum #1\expandafter \@firstoftwo
 \else \expandafter \@secondoftwo
 \fi
}%
\providecommand \@ifx [1]{%
 \ifx #1\expandafter \@firstoftwo
 \else \expandafter \@secondoftwo
 \fi
}%
\providecommand \natexlab [1]{#1}%
\providecommand \enquote  [1]{``#1''}%
\providecommand \bibnamefont  [1]{#1}%
\providecommand \bibfnamefont [1]{#1}%
\providecommand \citenamefont [1]{#1}%
\providecommand \href@noop [0]{\@secondoftwo}%
\providecommand \href [0]{\begingroup \@sanitize@url \@href}%
\providecommand \@href[1]{\@@startlink{#1}\@@href}%
\providecommand \@@href[1]{\endgroup#1\@@endlink}%
\providecommand \@sanitize@url [0]{\catcode `\\12\catcode `\$12\catcode
  `\&12\catcode `\#12\catcode `\^12\catcode `\_12\catcode `\%12\relax}%
\providecommand \@@startlink[1]{}%
\providecommand \@@endlink[0]{}%
\providecommand \url  [0]{\begingroup\@sanitize@url \@url }%
\providecommand \@url [1]{\endgroup\@href {#1}{\urlprefix }}%
\providecommand \urlprefix  [0]{URL }%
\providecommand \Eprint [0]{\href }%
\providecommand \doibase [0]{http://dx.doi.org/}%
\providecommand \selectlanguage [0]{\@gobble}%
\providecommand \bibinfo  [0]{\@secondoftwo}%
\providecommand \bibfield  [0]{\@secondoftwo}%
\providecommand \translation [1]{[#1]}%
\providecommand \BibitemOpen [0]{}%
\providecommand \bibitemStop [0]{}%
\providecommand \bibitemNoStop [0]{.\EOS\space}%
\providecommand \EOS [0]{\spacefactor3000\relax}%
\providecommand \BibitemShut  [1]{\csname bibitem#1\endcsname}%
\let\auto@bib@innerbib\@empty
\bibitem [{\citenamefont {Seminara}(2010)}]{seminara2010fluvial}%
  \BibitemOpen
  \bibfield  {author} {\bibinfo {author} {\bibfnamefont {G.}~\bibnamefont
  {Seminara}},\ }\href@noop {} {\bibfield  {journal} {\bibinfo  {journal}
  {Annual Review of Fluid Mechanics}\ }\textbf {\bibinfo {volume} {42}},\
  \bibinfo {pages} {43} (\bibinfo {year} {2010})}\BibitemShut {NoStop}%
\bibitem [{\citenamefont {Charru}\ \emph {et~al.}(2013)\citenamefont {Charru},
  \citenamefont {Andreotti},\ and\ \citenamefont {Claudin}}]{charru2013sand}%
  \BibitemOpen
  \bibfield  {author} {\bibinfo {author} {\bibfnamefont {F.}~\bibnamefont
  {Charru}}, \bibinfo {author} {\bibfnamefont {B.}~\bibnamefont {Andreotti}}, \
  and\ \bibinfo {author} {\bibfnamefont {P.}~\bibnamefont {Claudin}},\
  }\href@noop {} {\bibfield  {journal} {\bibinfo  {journal} {Annual Review of
  Fluid Mechanics}\ }\textbf {\bibinfo {volume} {45}},\ \bibinfo {pages} {469}
  (\bibinfo {year} {2013})}\BibitemShut {NoStop}%
\bibitem [{\citenamefont {Ruth}\ \emph {et~al.}(1933)\citenamefont {Ruth},
  \citenamefont {Montillon},\ and\ \citenamefont {Montonna}}]{ruth1933studies}%
  \BibitemOpen
  \bibfield  {author} {\bibinfo {author} {\bibfnamefont {B.}~\bibnamefont
  {Ruth}}, \bibinfo {author} {\bibfnamefont {G.}~\bibnamefont {Montillon}}, \
  and\ \bibinfo {author} {\bibfnamefont {R.}~\bibnamefont {Montonna}},\
  }\href@noop {} {\bibfield  {journal} {\bibinfo  {journal} {Industrial \&
  Engineering Chemistry}\ }\textbf {\bibinfo {volume} {25}},\ \bibinfo {pages}
  {76} (\bibinfo {year} {1933})}\BibitemShut {NoStop}%
\bibitem [{\citenamefont {Wyss}\ \emph {et~al.}(2006)\citenamefont {Wyss},
  \citenamefont {Blair}, \citenamefont {Morris}, \citenamefont {Stone},\ and\
  \citenamefont {Weitz}}]{wyss2006mechanism}%
  \BibitemOpen
  \bibfield  {author} {\bibinfo {author} {\bibfnamefont {H.~M.}\ \bibnamefont
  {Wyss}}, \bibinfo {author} {\bibfnamefont {D.~L.}\ \bibnamefont {Blair}},
  \bibinfo {author} {\bibfnamefont {J.~F.}\ \bibnamefont {Morris}}, \bibinfo
  {author} {\bibfnamefont {H.~A.}\ \bibnamefont {Stone}}, \ and\ \bibinfo
  {author} {\bibfnamefont {D.~A.}\ \bibnamefont {Weitz}},\ }\href@noop {}
  {\bibfield  {journal} {\bibinfo  {journal} {Physical review E}\ }\textbf
  {\bibinfo {volume} {74}},\ \bibinfo {pages} {061402} (\bibinfo {year}
  {2006})}\BibitemShut {NoStop}%
\bibitem [{\citenamefont {Dressaire}\ and\ \citenamefont
  {Sauret}(2017)}]{dressaire2017clogging}%
  \BibitemOpen
  \bibfield  {author} {\bibinfo {author} {\bibfnamefont {E.}~\bibnamefont
  {Dressaire}}\ and\ \bibinfo {author} {\bibfnamefont {A.}~\bibnamefont
  {Sauret}},\ }\href@noop {} {\bibfield  {journal} {\bibinfo  {journal} {Soft
  Matter}\ }\textbf {\bibinfo {volume} {13}},\ \bibinfo {pages} {37} (\bibinfo
  {year} {2017})}\BibitemShut {NoStop}%
\bibitem [{\citenamefont {Werner}\ \emph {et~al.}(2007)\citenamefont {Werner},
  \citenamefont {Jones}, \citenamefont {Paterson}, \citenamefont {Archer},\
  and\ \citenamefont {Pearce}}]{werner2007air}%
  \BibitemOpen
  \bibfield  {author} {\bibinfo {author} {\bibfnamefont {S.~R.}\ \bibnamefont
  {Werner}}, \bibinfo {author} {\bibfnamefont {J.~R.}\ \bibnamefont {Jones}},
  \bibinfo {author} {\bibfnamefont {A.~H.}\ \bibnamefont {Paterson}}, \bibinfo
  {author} {\bibfnamefont {R.~H.}\ \bibnamefont {Archer}}, \ and\ \bibinfo
  {author} {\bibfnamefont {D.~L.}\ \bibnamefont {Pearce}},\ }\href@noop {}
  {\bibfield  {journal} {\bibinfo  {journal} {Powder Technology}\ }\textbf
  {\bibinfo {volume} {171}},\ \bibinfo {pages} {25} (\bibinfo {year}
  {2007})}\BibitemShut {NoStop}%
\bibitem [{\citenamefont {Iveson}\ \emph {et~al.}(2001)\citenamefont {Iveson},
  \citenamefont {Litster}, \citenamefont {Hapgood},\ and\ \citenamefont
  {Ennis}}]{iveson2001nucleation}%
  \BibitemOpen
  \bibfield  {author} {\bibinfo {author} {\bibfnamefont {S.~M.}\ \bibnamefont
  {Iveson}}, \bibinfo {author} {\bibfnamefont {J.~D.}\ \bibnamefont {Litster}},
  \bibinfo {author} {\bibfnamefont {K.}~\bibnamefont {Hapgood}}, \ and\
  \bibinfo {author} {\bibfnamefont {B.~J.}\ \bibnamefont {Ennis}},\ }\href@noop
  {} {\bibfield  {journal} {\bibinfo  {journal} {Powder technology}\ }\textbf
  {\bibinfo {volume} {117}},\ \bibinfo {pages} {3} (\bibinfo {year}
  {2001})}\BibitemShut {NoStop}%
\bibitem [{\citenamefont {Deen}\ \emph {et~al.}(2007)\citenamefont {Deen},
  \citenamefont {Annaland}, \citenamefont {Van~der Hoef},\ and\ \citenamefont
  {Kuipers}}]{deen2007review}%
  \BibitemOpen
  \bibfield  {author} {\bibinfo {author} {\bibfnamefont {N.}~\bibnamefont
  {Deen}}, \bibinfo {author} {\bibfnamefont {M.~V.~S.}\ \bibnamefont
  {Annaland}}, \bibinfo {author} {\bibfnamefont {M.}~\bibnamefont {Van~der
  Hoef}}, \ and\ \bibinfo {author} {\bibfnamefont {J.}~\bibnamefont
  {Kuipers}},\ }\href@noop {} {\bibfield  {journal} {\bibinfo  {journal}
  {Chemical Engineering Science}\ }\textbf {\bibinfo {volume} {62}},\ \bibinfo
  {pages} {28} (\bibinfo {year} {2007})}\BibitemShut {NoStop}%
\bibitem [{\citenamefont {Varas}\ \emph {et~al.}(2011)\citenamefont {Varas},
  \citenamefont {Vidal},\ and\ \citenamefont
  {G{\'e}minard}}]{varas2011venting}%
  \BibitemOpen
  \bibfield  {author} {\bibinfo {author} {\bibfnamefont {G.}~\bibnamefont
  {Varas}}, \bibinfo {author} {\bibfnamefont {V.}~\bibnamefont {Vidal}}, \ and\
  \bibinfo {author} {\bibfnamefont {J.-C.}\ \bibnamefont {G{\'e}minard}},\
  }\href@noop {} {\bibfield  {journal} {\bibinfo  {journal} {Physical Review
  E}\ }\textbf {\bibinfo {volume} {83}},\ \bibinfo {pages} {011302} (\bibinfo
  {year} {2011})}\BibitemShut {NoStop}%
\bibitem [{\citenamefont {Ramos}\ \emph {et~al.}(2015)\citenamefont {Ramos},
  \citenamefont {Varas}, \citenamefont {G{\'e}minard},\ and\ \citenamefont
  {Vidal}}]{ramos2015gas}%
  \BibitemOpen
  \bibfield  {author} {\bibinfo {author} {\bibfnamefont {G.}~\bibnamefont
  {Ramos}}, \bibinfo {author} {\bibfnamefont {G.}~\bibnamefont {Varas}},
  \bibinfo {author} {\bibfnamefont {J.-C.}\ \bibnamefont {G{\'e}minard}}, \
  and\ \bibinfo {author} {\bibfnamefont {V.}~\bibnamefont {Vidal}},\
  }\href@noop {} {\bibfield  {journal} {\bibinfo  {journal} {Physical Review
  E}\ }\textbf {\bibinfo {volume} {92}},\ \bibinfo {pages} {062210} (\bibinfo
  {year} {2015})}\BibitemShut {NoStop}%
\bibitem [{\citenamefont {Carazzo}\ and\ \citenamefont
  {Jellinek}(2013)}]{carazzo2013particle}%
  \BibitemOpen
  \bibfield  {author} {\bibinfo {author} {\bibfnamefont {G.}~\bibnamefont
  {Carazzo}}\ and\ \bibinfo {author} {\bibfnamefont {A.}~\bibnamefont
  {Jellinek}},\ }\href@noop {} {\bibfield  {journal} {\bibinfo  {journal}
  {Journal of Geophysical Research: Solid Earth}\ }\textbf {\bibinfo {volume}
  {118}},\ \bibinfo {pages} {1420} (\bibinfo {year} {2013})}\BibitemShut
  {NoStop}%
\bibitem [{\citenamefont {Munro}\ \emph {et~al.}(2009)\citenamefont {Munro},
  \citenamefont {Bethke},\ and\ \citenamefont {Dalziel}}]{munro2009sediment}%
  \BibitemOpen
  \bibfield  {author} {\bibinfo {author} {\bibfnamefont {R.~J.}\ \bibnamefont
  {Munro}}, \bibinfo {author} {\bibfnamefont {N.}~\bibnamefont {Bethke}}, \
  and\ \bibinfo {author} {\bibfnamefont {S.}~\bibnamefont {Dalziel}},\
  }\href@noop {} {\bibfield  {journal} {\bibinfo  {journal} {Physics of
  Fluids}\ }\textbf {\bibinfo {volume} {21}},\ \bibinfo {pages} {046601}
  (\bibinfo {year} {2009})}\BibitemShut {NoStop}%
\bibitem [{\citenamefont {Bethke}\ and\ \citenamefont
  {Dalziel}(2012)}]{bethke2012resuspension}%
  \BibitemOpen
  \bibfield  {author} {\bibinfo {author} {\bibfnamefont {N.}~\bibnamefont
  {Bethke}}\ and\ \bibinfo {author} {\bibfnamefont {S.~B.}\ \bibnamefont
  {Dalziel}},\ }\href@noop {} {\bibfield  {journal} {\bibinfo  {journal}
  {Physics of Fluids}\ }\textbf {\bibinfo {volume} {24}},\ \bibinfo {pages}
  {063301} (\bibinfo {year} {2012})}\BibitemShut {NoStop}%
\bibitem [{\citenamefont {Yoshida}\ \emph {et~al.}(2012)\citenamefont
  {Yoshida}, \citenamefont {Masuda}, \citenamefont {Ito}, \citenamefont
  {Furuya},\ and\ \citenamefont {Sano}}]{yoshida2012collision}%
  \BibitemOpen
  \bibfield  {author} {\bibinfo {author} {\bibfnamefont {J.}~\bibnamefont
  {Yoshida}}, \bibinfo {author} {\bibfnamefont {N.}~\bibnamefont {Masuda}},
  \bibinfo {author} {\bibfnamefont {B.}~\bibnamefont {Ito}}, \bibinfo {author}
  {\bibfnamefont {T.}~\bibnamefont {Furuya}}, \ and\ \bibinfo {author}
  {\bibfnamefont {O.}~\bibnamefont {Sano}},\ }\href@noop {} {\bibfield
  {journal} {\bibinfo  {journal} {Fluid Dynamics Research}\ }\textbf {\bibinfo
  {volume} {44}},\ \bibinfo {pages} {015502} (\bibinfo {year}
  {2012})}\BibitemShut {NoStop}%
\bibitem [{\citenamefont {Masuda}\ \emph {et~al.}(2012)\citenamefont {Masuda},
  \citenamefont {Yoshida}, \citenamefont {Ito}, \citenamefont {Furuya},\ and\
  \citenamefont {Sano}}]{masuda2012collision}%
  \BibitemOpen
  \bibfield  {author} {\bibinfo {author} {\bibfnamefont {N.}~\bibnamefont
  {Masuda}}, \bibinfo {author} {\bibfnamefont {J.}~\bibnamefont {Yoshida}},
  \bibinfo {author} {\bibfnamefont {B.}~\bibnamefont {Ito}}, \bibinfo {author}
  {\bibfnamefont {T.}~\bibnamefont {Furuya}}, \ and\ \bibinfo {author}
  {\bibfnamefont {O.}~\bibnamefont {Sano}},\ }\href@noop {} {\bibfield
  {journal} {\bibinfo  {journal} {Fluid Dynamics Research}\ }\textbf {\bibinfo
  {volume} {44}},\ \bibinfo {pages} {015501} (\bibinfo {year}
  {2012})}\BibitemShut {NoStop}%
\bibitem [{\citenamefont {Badr}\ \emph {et~al.}(2014)\citenamefont {Badr},
  \citenamefont {Gauthier},\ and\ \citenamefont {Gondret}}]{badr2014erosion}%
  \BibitemOpen
  \bibfield  {author} {\bibinfo {author} {\bibfnamefont {S.}~\bibnamefont
  {Badr}}, \bibinfo {author} {\bibfnamefont {G.}~\bibnamefont {Gauthier}}, \
  and\ \bibinfo {author} {\bibfnamefont {P.}~\bibnamefont {Gondret}},\
  }\href@noop {} {\bibfield  {journal} {\bibinfo  {journal} {Physics of
  Fluids}\ }\textbf {\bibinfo {volume} {26}},\ \bibinfo {pages} {023302}
  (\bibinfo {year} {2014})}\BibitemShut {NoStop}%
\bibitem [{\citenamefont {Sutherland}\ and\ \citenamefont
  {Dalziel}(2014)}]{sutherland2014bedload}%
  \BibitemOpen
  \bibfield  {author} {\bibinfo {author} {\bibfnamefont {B.~R.}\ \bibnamefont
  {Sutherland}}\ and\ \bibinfo {author} {\bibfnamefont {S.}~\bibnamefont
  {Dalziel}},\ }\href@noop {} {\bibfield  {journal} {\bibinfo  {journal}
  {Physics of Fluids}\ }\textbf {\bibinfo {volume} {26}},\ \bibinfo {pages}
  {035103} (\bibinfo {year} {2014})}\BibitemShut {NoStop}%
\bibitem [{\citenamefont {Charru}\ \emph {et~al.}(2004)\citenamefont {Charru},
  \citenamefont {Mouilleron},\ and\ \citenamefont {Eiff}}]{charru2004erosion}%
  \BibitemOpen
  \bibfield  {author} {\bibinfo {author} {\bibfnamefont {F.}~\bibnamefont
  {Charru}}, \bibinfo {author} {\bibfnamefont {H.}~\bibnamefont {Mouilleron}},
  \ and\ \bibinfo {author} {\bibfnamefont {O.}~\bibnamefont {Eiff}},\
  }\href@noop {} {\bibfield  {journal} {\bibinfo  {journal} {Journal of Fluid
  Mechanics}\ }\textbf {\bibinfo {volume} {519}},\ \bibinfo {pages} {55}
  (\bibinfo {year} {2004})}\BibitemShut {NoStop}%
\bibitem [{\citenamefont {Hong}\ \emph {et~al.}(2015)\citenamefont {Hong},
  \citenamefont {Tao},\ and\ \citenamefont {Kudrolli}}]{hong2015onset}%
  \BibitemOpen
  \bibfield  {author} {\bibinfo {author} {\bibfnamefont {A.}~\bibnamefont
  {Hong}}, \bibinfo {author} {\bibfnamefont {M.}~\bibnamefont {Tao}}, \ and\
  \bibinfo {author} {\bibfnamefont {A.}~\bibnamefont {Kudrolli}},\ }\href@noop
  {} {\bibfield  {journal} {\bibinfo  {journal} {Physics of Fluids}\ }\textbf
  {\bibinfo {volume} {27}},\ \bibinfo {pages} {013301} (\bibinfo {year}
  {2015})}\BibitemShut {NoStop}%
\bibitem [{\citenamefont {Solomatov}\ \emph {et~al.}(1993)\citenamefont
  {Solomatov}, \citenamefont {Olson},\ and\ \citenamefont
  {Stevenson}}]{solomatov1993entrainment}%
  \BibitemOpen
  \bibfield  {author} {\bibinfo {author} {\bibfnamefont {V.~S.}\ \bibnamefont
  {Solomatov}}, \bibinfo {author} {\bibfnamefont {P.}~\bibnamefont {Olson}}, \
  and\ \bibinfo {author} {\bibfnamefont {D.~J.}\ \bibnamefont {Stevenson}},\
  }\href@noop {} {\bibfield  {journal} {\bibinfo  {journal} {Earth and
  planetary science letters}\ }\textbf {\bibinfo {volume} {120}},\ \bibinfo
  {pages} {387} (\bibinfo {year} {1993})}\BibitemShut {NoStop}%
\bibitem [{\citenamefont {Solomatov}\ and\ \citenamefont
  {Stevenson}(1993)}]{solomatov1993suspension}%
  \BibitemOpen
  \bibfield  {author} {\bibinfo {author} {\bibfnamefont {V.~S.}\ \bibnamefont
  {Solomatov}}\ and\ \bibinfo {author} {\bibfnamefont {D.~J.}\ \bibnamefont
  {Stevenson}},\ }\href@noop {} {\bibfield  {journal} {\bibinfo  {journal}
  {Journal of Geophysical Research: Planets}\ }\textbf {\bibinfo {volume}
  {98}},\ \bibinfo {pages} {5375} (\bibinfo {year} {1993})}\BibitemShut
  {NoStop}%
\bibitem [{\citenamefont {Martin}\ and\ \citenamefont
  {Nokes}(1988)}]{martin1988crystal}%
  \BibitemOpen
  \bibfield  {author} {\bibinfo {author} {\bibfnamefont {D.}~\bibnamefont
  {Martin}}\ and\ \bibinfo {author} {\bibfnamefont {R.}~\bibnamefont {Nokes}},\
  }\href@noop {} {\bibfield  {journal} {\bibinfo  {journal} {Nature}\ }\textbf
  {\bibinfo {volume} {332}},\ \bibinfo {pages} {534} (\bibinfo {year}
  {1988})}\BibitemShut {NoStop}%
\bibitem [{\citenamefont {Boyer}(2011)}]{boyer2011suspensions}%
  \BibitemOpen
  \bibfield  {author} {\bibinfo {author} {\bibfnamefont {F.}~\bibnamefont
  {Boyer}},\ }\emph {\bibinfo {title} {Suspensions concentr{\'e}es:
  Exp{\'e}riences originales de rh{\'e}ologie}},\ \href@noop {} {Ph.D.
  thesis},\ \bibinfo  {school} {Aix Marseille 1} (\bibinfo {year}
  {2011})\BibitemShut {NoStop}%
\bibitem [{\citenamefont {Hartline}\ and\ \citenamefont
  {Lister}(1977)}]{Hartline1977}%
  \BibitemOpen
  \bibfield  {author} {\bibinfo {author} {\bibfnamefont {B.~K.}\ \bibnamefont
  {Hartline}}\ and\ \bibinfo {author} {\bibfnamefont {C.~R.~B.}\ \bibnamefont
  {Lister}},\ }\href@noop {} {\bibfield  {journal} {\bibinfo  {journal}
  {Journal of Fluid Mechanics}\ }\textbf {\bibinfo {volume} {79}},\ \bibinfo
  {pages} {379} (\bibinfo {year} {1977})}\BibitemShut {NoStop}%
\bibitem [{\citenamefont {Davaille}(1999)}]{Davaille1999}%
  \BibitemOpen
  \bibfield  {author} {\bibinfo {author} {\bibfnamefont {A.}~\bibnamefont
  {Davaille}},\ }\href {\doibase 10.1017/S0022112098003322} {\emph {\bibinfo
  {title} {Journal of Fluid Mechanics}}},\ Vol.\ \bibinfo {volume} {379}\
  (\bibinfo {year} {1999})\ pp.\ \bibinfo {pages} {223--253}\BibitemShut
  {NoStop}%
\bibitem [{\citenamefont {{Le Bars}}\ and\ \citenamefont
  {Davaille}(2002)}]{LeBars2002}%
  \BibitemOpen
  \bibfield  {author} {\bibinfo {author} {\bibfnamefont {M.}~\bibnamefont {{Le
  Bars}}}\ and\ \bibinfo {author} {\bibfnamefont {A.}~\bibnamefont
  {Davaille}},\ }\href {\doibase 10.1017/S0022112002001878} {\bibfield
  {journal} {\bibinfo  {journal} {Journal of Fluid Mechanics}\ }\textbf
  {\bibinfo {volume} {471}},\ \bibinfo {pages} {339} (\bibinfo {year}
  {2002})}\BibitemShut {NoStop}%
\bibitem [{\citenamefont {{Le Bars}}\ and\ \citenamefont
  {Davaille}(2004)}]{LeBars2004}%
  \BibitemOpen
  \bibfield  {author} {\bibinfo {author} {\bibfnamefont {M.}~\bibnamefont {{Le
  Bars}}}\ and\ \bibinfo {author} {\bibfnamefont {A.}~\bibnamefont
  {Davaille}},\ }\href@noop {} {\bibfield  {journal} {\bibinfo  {journal}
  {Journal of Fluid Mechanics}\ }\textbf {\bibinfo {volume} {499}},\ \bibinfo
  {pages} {75} (\bibinfo {year} {2004})}\BibitemShut {NoStop}%
   \bibitem{polystyrene} A. Abe and D. R. Bloch (1989). \textit{Polymer handbook} (Vol. 7). J. Brandrup, E. H. Immergut, \& E. A. Grulke (Eds.). New York etc: Wiley.
\end{thebibliography}
\end{document}